\documentclass[12pt]{article}
\usepackage{epsf}       %this style file is for embedding postscript file
\usepackage{graphicx}       %this style file is for embedding postscript file

\begin{document}
\title{The NA48 Liquid Krypton Calorimeter \\
Description and Performances}
\author{Jos\'e OCARIZ\thanks{
e-mail:{\tt jose.ocariz@cern.ch}} \\
Laboratoire de l'Acc\'el\'erateur Lin\'eaire, \\ 
IN2P3-CNRS and Universit\'e Paris-Sud \\
B.P. 34, 
91898 Orsay Cedex, France\\
for the 
NA48 Collaboration
\thanks{The NA48 Collaboration is:
Cagliari, Cambridge, CERN, Dubna,
Edinburgh, Ferrara, Firenze,
Mainz, Orsay, Perugia, Pisa, Saclay, Siegen, Torino, Vienna, Warsaw.}
}
\maketitle
\begin{abstract}
The NA48 experiment at CERN\cite{ref1} aims at making a precision 
study of direct
CP violation in the neutral kaons, by measuring
$Re\left( \frac{\varepsilon '}{\varepsilon}\right) $
with an accuracy better than $0.02\%$. 
To achieve this goal, the experiment requires a neutral detector
with fast response, high efficiency in a high-rate environment
($\sim$ {\tt MHz}), long-term stability, sub-nanosecond time resolution,
millimetric space  precision, and an excellent energy
resolution ($1\%$) in the $5\to100$ {\tt GeV} range. 
To achieve these
performances, a quasi-homogeneous Liquid Krypton calorimeter
has been chosen, designed with a projective tower geometry,
high transversal segmentation, and fast digital readout. The
calorimeter was operative during the '97 data taking period,
its performances were thoroughly studied, and found to be in
agreement with design requirements. A detector description and
performances analysis are here presented.
\end{abstract}
\newpage
\section{Introduction}
The NA48 collaboration attempts at measuring the ${\cal CP}$-violating
parameter $Re\left( \frac{\varepsilon '}{\varepsilon}\right) $
with an accuracy $\simeq 2\times 10^{-4}$. A non--zero value of $\varepsilon'$
 would constitute an experimental evidence
of direct ${\cal CP}$ violation.  This quantity
can be extracted out of the experimentally
observable ``double ratio'' $R$,
\begin{eqnarray}
\nonumber
R&=&\left| \frac{\eta^{00}}{\eta^{+-}}\right|^2
=\frac{\Gamma \left( K_L\to \pi^0 \pi^0 \right)}
{\Gamma \left( K_S\to \pi^0 \pi^0 \right)}
\ / \
\frac{\Gamma \left( K_L\to \pi^+ \pi^- \right)}
{\Gamma \left( K_S\to \pi^+ \pi^- \right)}
\\ 
&=&1-6 {\it Re}\left( \frac{\varepsilon'}{\varepsilon}\right)
+{\cal O}\left(\frac{\varepsilon'}{\varepsilon}\right)^2
\end{eqnarray}
The statistical error being dominated by the number of $K_L\to2\pi^0$
decays, a few million of them are to be identified. In a similar way,
all systematic uncertainties must be minimised. The NA48 detector has
beed designed to cope with these two requirements.
\section{The NA48 Detector}
\begin{itemize}
\item two simultaneous, quasi-collinear  $K_L$ and
 $K_S$ beams,
\item a {\it tagger} to identify decays from each beam,
\item a {\it magnetic spectrometer} to measure the charged
modes $K_{L,S}\to \pi^+\pi^-$,
\item a  {\it liquid krypton calorimeter} for the neutral modes $K_{L,S}\to 2\pi^0\to4\gamma$,
\item other subdetectors for background rejection
and redundant identification .
\end{itemize}
In order to keep a steady acquisition in a high-rate environment 
($\sim MHz$) during several years, the detectors are required to be
fast, efficient and stable. Moreover, ${\cal CP}$-violating
decay channels have to be identified in presence of important
backgrounds, e.g.
\begin{center}
\begin{tabular}{cc}
signal& dominant background
\\
\hline
 $K_L\to 2\pi^0$ & $K_L\to 3\pi^0$
\\
B.R. $\sim 0.09\%$ & B.R. $\sim 21.6 \%$
\\
\hline
$K_L\to \pi^+\pi^-$ & $K_L\to \pi e\nu_e$
\\
B.R.  $\sim 0.2\%$ & B.R.  $\sim 37.8 \%$
\end{tabular}
\end{center}
The analysis requires backgrounds to be rejected to
a $\ ^{<}_{\sim} \ 0.1\%$ level. This condition can be translated in terms of  calorimetric 
performances as:
\begin{center}
$\begin{array}{cccc}
\sigma_T & < & 500. \ {\tt ps} &
\\
\sigma_{X,Y} & < & 1. \ {\tt mm} &
\\
\sigma_E/E & ^{<}_{\sim} & 1. \% & \left( E_\gamma  
\sim 25 \ {\tt GeV}\right) 
\\
{\tt linearity} & \simeq & {\tt few} \ 10^{-3} & (5 \to 100 \ {\tt GeV})
\\ 
\end{array}$
\end{center}
\section{The Liquid Krypton Calorimeter}
NA48 uses a quasi-homogeneous calorimeter, made up of some  $10 \ {\tt m}^3$
of liquid krypton. In order to
ensure a good transverse granularity, given the
krypton  $6.1$ {\tt cm} Moli\`ere radius,
the calorimeter is segmentated into $13212$ cells of area $2\times 2$ ${\tt cm}^2$.
The calorimeter depth is $1.25$ {\tt m},
which corresponds to some $27$ radiation lengths ($X_0\simeq 4.7$ {\tt cm}).

The calorimeter is built following a projective geometry, pointing towards
the Kaon fiducial decay region (about $114$ {\tt m} upstream). In order to minimise energy losses for showers
developing close to the anodes,  ribbon electrodes follow a $48$ {\tt mrad} zig-zag 
along the projectivity lines. To ensure a 
gap uniformity within $0.4 \%$, distance between electrodes
is kept constant by means of regular spacers.
\subsection{The LKr electronics readout}
In order to minimise noise, the induced triangular signal in the cell 
electrodes
(drift time $3$ $\mu{\tt s}$) is read by preamplifiers\cite{ref2} 
(PA) inside the krypton. It is then
differentiated by transceivers\cite{ref3} (time constant $RC=10 
{\tt ns}$) and sent 
through twisted cables to the Calorimeter Pipeline Digitisers\cite{ref4} (CPD),
for final shaping. The CPD signal generates a sharp peak, with a $70$ {\tt ns} FWHM,
that is processed by means of a optimal filtering algorithm to evaluate pulse
height and extract energy and time.

Asynchronous digital sampling is done every $25$ {\tt ns}, with a
$4$ gain, $10$ ADC bit dynamic range, covering an energy interval
from  $3.5$ {\tt MeV}
up to $50$ {\tt GeV} per channel. Gain dispersion
being the order of $3\%$, a calibration signal is sent at the PA level and 
used for 
gain equalisation and offset determination. This calibration procedure,
done in parallel with physics data acquisition,
shows that gain stability is maintained  within a $0.1\%$ level. 
Offset variations (typically of the order of $0.1$  {\tt ADC}/ $^0C$) 
are permanently monitored.
 \begin{figure}[ht]
\centerline{\includegraphics[width=10cm]{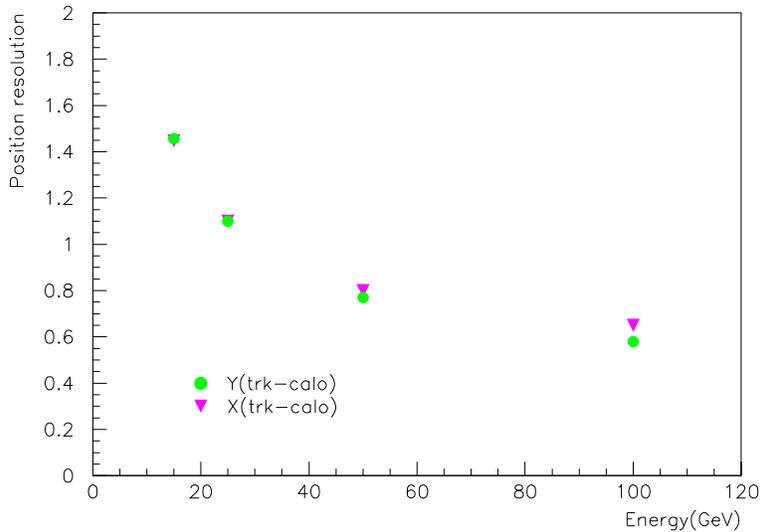}}
\caption{
LKr and spectrometer position measurement for electrons.
Vertical scale is in {\tt mm}.
}
\label{fig1}
\end{figure}
\section{LKr performances}
LKr electronics was completely setup for the 1997 data-taking period. High voltage 
was set at $1500$ Volts to prevent failure
of a few blocking capacitors; also, one LKr column (out of $64$) disconnected from high
voltage during cooling, generating a $\sim 15\%$ acceptance 
loss for the neutral modes. Some
$650 \ 000$ $K_L\to 2\pi^0$ decays were successfully collected by the end of the
'97 run, and the physics data was used to evaluate the LKr performances.

\subsection{LKr time and position resolutions with electrons}
SPS electrons at fixed energies ($15$, $25$, $50$, and $100$ {\tt GeV}) were used to
evaluate the position and time performances.

\begin{figure}[ht]
\centerline{\includegraphics[width=10cm]{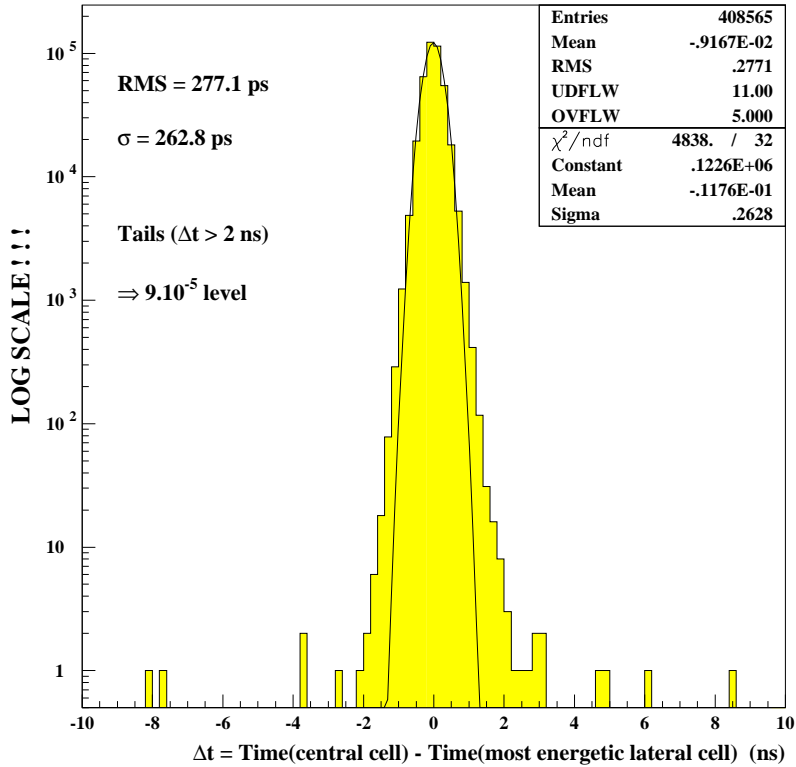} }
\caption{
LKr time difference for neighbour cells.
}
\label{fig2}
\end{figure}

Figure \ref{fig1} shows 
the position resolution as a function of electron energy. Above $25$ {\tt GeV},
both horizontal and vertical
projections are evaluated with a resolution better than $1$ {\tt mm}.
Intrinsic time performances are defined as the time difference for the
most energetic neighbour cells in an electron shower. 
Figure
\ref{fig2} shows a time resolution of
$260$ {\tt ps}, while small inefficiencies (i.e. time differences
greater than $2$ {\tt ns}) are below $10^{-4}$,  and were shown to be linked to
accidentals\cite{ref5}.
\subsection{LKr energy performances with $K_{e3}$}
$K_{e3}$ decays ($K_L\to \pi^{\pm}e^{\mp}\nu$) are used to compare  electron 
momentum $p$, as measured by the magnetic spectrometer, with  LKr
electron shower energy $E$. This allows to test calibration, energy resolution and
linearity.
 \begin{figure}[ht]
\centerline{\includegraphics[width=10cm]{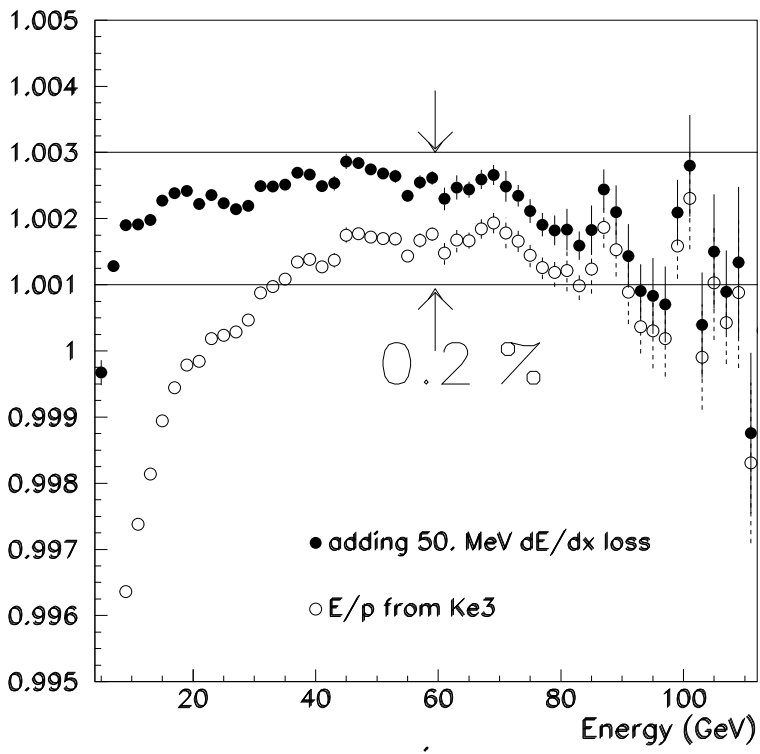} }
\caption{
LKr linearity
}
\label{fig3}
\end{figure}
A GEANT simulation shows that $K_{e3}$ electrons lose in average  some $50$ {\tt MeV}
before entering the cryostat. Figure \ref{fig3} shows the $E/P$
variation with energy, before and after correcting for this $dE/dx$
energy loss. LKr energy
 response is  linear within $0.2\%$
in the $5\to 100$ $GeV$ range.
 \begin{figure}[ht]
\centerline{\includegraphics[width=10cm]{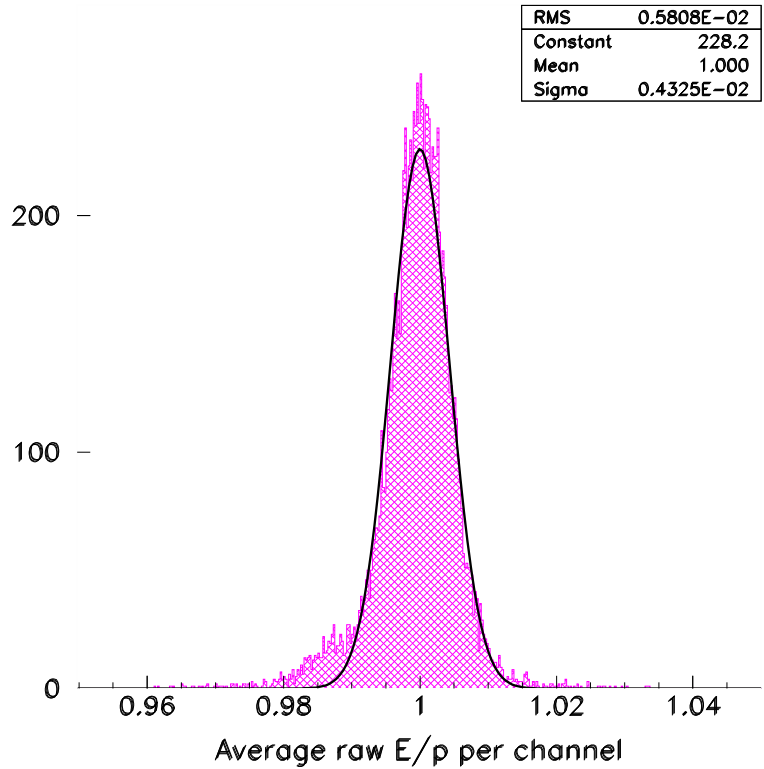} }
\caption{
LKr cell-to-cell energy uniformity
}
\label{fig4}
\end{figure}
Cell-to-cell uniformity (i.e. average $E/p$ read by a single channel) 
is uniform within  $0.43 \%$, as shown in figure \ref{fig4}. The remaining variations
were shown to be partially linked with uncertainties in the calibration
electronics, and an iterative procedure with $K_{e3}$ electrons was used to evaluate
a set of cell energy factors, and optimise cell energy readout uniformity.
 \begin{figure}[ht]
\centerline{\includegraphics[width=10cm]{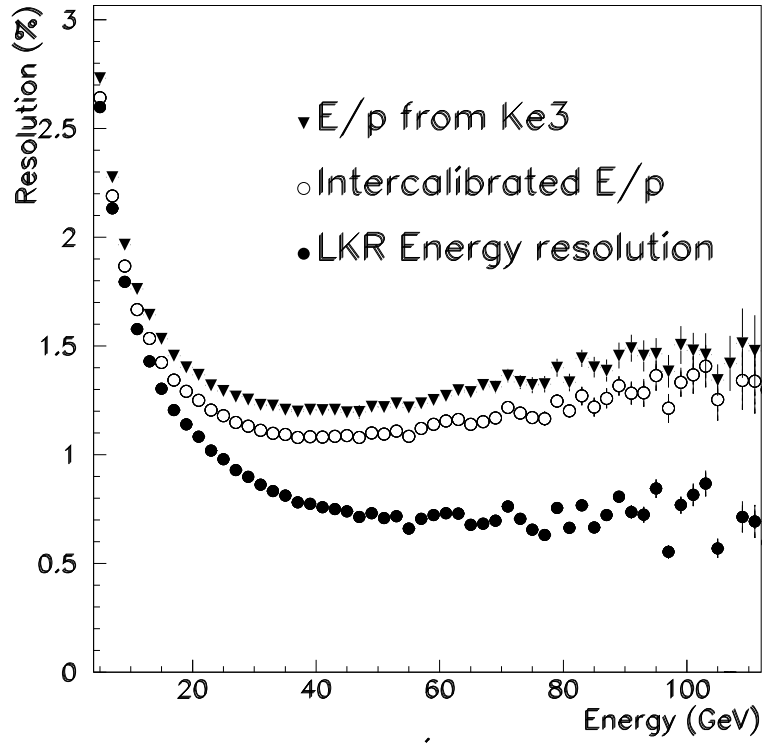} }
\caption{
LKr energy resolution
}
\label{fig5}
\end{figure}
Magnetic spectrometer measurement of momentum worsening
with energy, its contribution has to be substracted from the $E/p$ resolution
to obtain the pure LKr energy performance. From figure \ref{fig5}, the
LKr energy resolution is evaluated as
\begin{eqnarray}
\frac{\sigma\left( E\right)}{E}=
\frac{3.5\%}{\sqrt{E}}\oplus \frac{110 \ MeV}{E}\oplus 0.6 \%
\end{eqnarray}
For the neutral modes detection the LKr performances yield a $1.1$ {\tt MeV} $\pi^0$
mass 
resolution. After weighting $K_L$ events, the remaining 
 $K_L\to 3\pi^0$ background under
the signal is reduced to $^{<}_{\sim} \ 0.1\%$.

\section{LKr in 1998}
The blocking capacitors were replaced during the '97-'98 winter shutdown, 
and '98 data-taking was achieved successfully. The
LKr was safely operated at $3000$ Volts, obtaining a $20\%$ signal
gain and a corresponding resolution improvement.

About $2\times 10^{6}$ $K_L\to 2\pi^0$ events were added to the NA48 statistics.
\section{Conclusions}
LKr performances were tested with physics data during the '97 data-taking
period. Its performances were shown to be within design requirements; in
particular, background to the neutral modes was rejected up to a
residual $^<_\sim$ $0.1\%$ level.

\end{document}